# Electronic and magnetic properties of epitaxial SrRhO$_3$ films

John Nichols[1], Simuck F. Yuk[1], Changhee Sohn[1], Hyoungjeen Jeen[1,2], John W. Freeland[3], Valentino R. Cooper[1], and Ho Nyung Lee[1]

[1]Materials Science and Technology Division, Oak Ridge National Laboratory, Oak Ridge, TN, 37831, USA
[2]Department of Physics, Pusan National University, Busan Korea 46241
[3]Advanced Photon Source, Argonne National Laboratory, Argonne, IL, 60439, USA

Abstract

Strong interplay of fundamental order parameters in complex oxides are known to give rise to exotic physical phenomena. The 4$d$ transition metal oxide SrRhO$_3$ has generated much interest, but advances have been hindered by difficulties in preparing single crystalline phases. Here, we have epitaxially stabilized high quality single crystalline SrRhO$_3$ films and investigated their structural, electronic, and magnetic properties. We determine that their properties significantly differ from the paramagnetic metallic ground state that governs bulk samples and are strongly related to rotations of the RhO$_6$ octahedra.

The delicate balance between charge, spin, orbital, and lattice order parameters has proven to give rise to novel properties in many transition metal oxides (TMOs).[1,2] In particular, the onset of unconventional p-wave superconductivity in $Sr_2RuO_4$[3] was the first example of Type-II superconductivity in an oxide without copper and inspired much interest in other analogous 4$d$ TMOs such as rhodates and other ruthenate compounds.[4,5] However, the synthesis of many of these compounds proved to be a formidable task, often requiring high pressure and high temperature; which in turn limited experimental efforts on many of these compounds to a few experimental studies with polycrystalline samples.[5-10] Theoretical work predicted that the magnetic ground state of the orthorhombic perovskite $SrRhO_3$ is sensitive to its symmetry with the bulk orthorhombic structure forming a paramagnetic (PM) metallic state, while the high symmetry cubic perovskite structure would reveal a ferromagnetic metallic state.[11] In addition, $SrRhO_3$ has recently received attention for the potential to realize topologically protected states as a constituent material in artificial heterostructures and superlattices.[12,13] The potential of $SrRhO_3$ to realize such exotic ground states has inspired us to overcome the technical challenges associated with creating phase pure $SrRhO_3$ by synthesizing epitaxial thin films.

In this manuscript, we report on the synthesis and basic physical properties of $SrRhO_3$ films grown on (001) $SrTiO_3$ (STO) substrates by pulsed laser epitaxy (PLE). Epitaxial stabilization by PLE is advantageous as meta-stable phases that are difficult, if not impossible, to grow can be synthesized.[14-16] Here, the first single crystalline samples of the orthorhombic perovskite $SrRhO_3$ were synthesized. We grew films with a KrF excimer laser ($\lambda$ = 248nm) with laser fluence, substrate temperature, and oxygen partial pressure of 1.0 J/cm$^2$, 750 ºC, and 100 mTorr, respectively. The samples were characterized by x-ray diffraction (XRD), atomic force microscopy (AFM), *dc*-transport, scanning quantum interference device (SQUID) magnetometry, x-ray absorption spectroscopy (XAS), and spectroscopic ellipsometry (SE). We confirmed from XRD studies that our SrRhO$_3$ films were epitaxial and coherently strained. At low temperature, the electronic and magnetic properties strongly deviate from the PM metallic ground state of bulk crystals, while there is an abnormally strong hybridization between the O 2$p$ and Rh 4$d$ orbitals. Density functional theory (DFT) calculations were conducted to better understand this material. Polycrystalline SrRhO$_3$ realizes a orthorhombic perovskite structure with lattice parameters $a$ = 5.5394 Å, $b$ = 7.8539 Å, and $c$ = 5.5666 Å (Space group *Pnma*) that result in a pseudo-cubic lattice constant of 3.93 Å with RhO$_6$ octahedra that undergo $a^+b^-b^-$ distortion,[5] as illustrated in Fig. 1.

The quality of these SrRhO$_3$ samples is confirmed through AFM and XRD measurements. Topographic images of a SrTiO$_3$ substrate prior to and after deposition of a 12 nm thick SrRhO$_3$ film are shown in Fig. 1a and 1b, respectively. Note that after deposition of the film, step terrace features of the underlying substrate are clearly visible, and the RMS roughness of the surface is ~2.2 Å, indicating that the epitaxial film is of high quality. The phase purity and orientation of the films are confirmed by XRD measurements where a $\theta$-2$\theta$ scan is shown in Fig. 1c. Note that the even SrRhO$_3$ 00$l$ reflections are clearly visible while the odd 00$l$ reflections are expected to be less intense than the background and are thus not visible here. The positions of the SrRhO$_3$ peaks are consistent with a compressively strained perovskite film that is elongated along the $c$-axis. The rocking curves have a full width at half maximum less than 0.04° and the reciprocal space maps (data not shown) indicate the high crystallinity and coherent strain state of the SrRhO$_3$ films.

Interestingly, this system has a critical thickness (~ 15nm) below which phase pure samples can be synthesized. Attempts to grow thicker films yielded an impurity phase that is most probably due to the onset of strain relaxation at this thickness which then results in one of the lower symmetry phases of strontium rhodate being more energetically favorable. In addition, this observation suggests that the in-plane compressive strain simulates the pressure required to produce the meta-stable high pressure perovskite phase of $SrRhO_3$.

The electronic properties of these $SrRhO_3$ films were investigated by *dc*-transport measurements as shown in Fig. 2a. Notice that at high temperature, the sample exhibits a metallic behavior consistent with the bulk properties of $SrRhO_3$. However, at low temperature, there is a weak upturn in the resistivity. At high temperature, there is a very weak temperature dependence indicating that the resistivity saturates near the Mott-Ioffe-Regel limit,[17,18] however here this limit is roughly four times larger than the range of 100—200 μΩ·cm that is commonly reported in the literature.[19] Recall that the $SrRhO_3$ films are ~12 nm thick and thus the upturn in resistivity at low temperature is likely due to the finite thickness effect which is known to result in an insulating behavior in ultrathin films of other correlated metallic oxide films.[20-24] The inset shows the Hall measurement results at 2 K, which displays a linear response with a carrier concentration and mobility of $3.8 \times 10^{21}$ cm$^{-3}$ and 1.1 cm$^2$ V$^{-1}$ s$^{-1}$, respectively. The temperature dependent magnetoresistance (MR) is presented in Fig. 2b where there is a sign change at ~110K that will be discussed below and an upturn at low temperature likely due to increased scattering off defects. The temperature dependent magnetization obtained from SQUID measurements is shown in Fig. 2c, which reveals a subtle anomaly at the same temperature as the sign change in MR, suggesting there is a change in the magnetic structure here and that this change is not ferromagnetic in origin.

In order to obtain a greater understanding of the low temperature properties that clearly differ from those of bulk crystals, DFT calculations were performed to determine the density of states (DOS) and magnetic ground state of these SrRhO$_3$ films. The crystal structure was refined using the lattice parameters of the coherently strained SrRhO$_3$ films of $a = b = 3.905$ Å and $c = 4.043$ Å with no RhO$_6$ rotations (space group *Amm2*). The atomic positions were then allowed to relax to the bulk-like orthorhombic structure with octahedral rotations (space group *Pnma*) while maintaining the thin film lattice parameters. Within this framework, the DFT calculations predict a ferromagnetic metal ground state which is strongly inconsistent with the SQUID data in Fig. 2. However, DFT calculations for the *Amm2* space group with no RhO$_6$ rotations (a$^0$a$^0$a$^0$) indicate that a C-type antiferromagnetic (AFM) ground state is energetically favorable, and the calculated DOS of this state is illustrated in Fig. 3a. There are two implications of this result. First, it provides further evidence of the high sensitivity of the magnetic ground state of SrRhO$_3$ to octahedra distortions,[11] Secondly, the octahedra in these SrRhO$_3$ films grown on cubic STO are not rotated due to the substrate, which transfers its symmetry across the heterointerface (i.e. cubic materials grown on orthorhombic substrates being orthorhombic and vice versa), similar to that observed in other systems.[25,26] This result was quite surprising considering bulk SrRhO$_3$ is an orthorhombic perovskite. So to test this, we performed XRD reciprocal space mapping for the all variants of the STO 103 reflection by azimuthally rotating the sample (see Fig. 4) and observed no evidence of orthorhombic distortions (i.e. RhO$_6$ rotations) in this SrRhO$_3$ film which has a cubic-like structure that is quite consistent with conclusions discussed above. Combining these DFT calculations with the MR and SQUID data, we determine that SrRhO$_3$ likely is a C-type AFM with a Neel temperature of T$_N$ ~ 110 K. Supporting this interpretation, the DOS for the C-type AFM state indicates that despite a gap-like feature forming near the Fermi level (E$_F$), it does not fully open and the DOS at $E_F$ remains finite. This result is quite consistent with the weak rise in resistivity with decreasing temperature.

The electronic properties of SrRhO$_3$ films were further investigated through XAS and SE measurements as presented in Fig. 5. Since these spectra were collected at room temperature, we expect that the PM DOS (Fig. 3b) will be most representative of the data since we anticipate a PM ground state here. Note that the PM and C-type AF DOS are qualitatively similar with the major distinction between the two being a suppression of states near $E_F$ in the latter. The XAS measurements were performed on beamline 4-ID-C at the Advanced Photon Source of Argonne National Laboratory and were obtained at room temperature with an angle of incidence of ~25º while simultaneously monitoring both fluorescence (TFY) and electron yields (TEY) near the O *K* edge. Recall that the samples are quite thin (t ~ 12 nm) so the TFY signal is strongly influenced by the underlying STO substrate as evident in the strong similarities between TFY and STO data in Fig. 1a. However even though TEY is known to be a surface sensitive probe, the similarities between both TFY and TEY at low energy indicates that TEY well represents the properties of the bulk of the film and is also shown in Fig. 5b. It is important to note that this data is somewhat distinct from similar measurements on polycrystalline SrRhO$_3$,[27] which also supports the claim that the physical properties of SrRhO$_3$ are highly sensitive to the local RhO$_6$ environment. First note that there is a small peak at ~523 eV that is due to the Rh *M*-edge and is too weak to be observed in TFY whereas all other features are due to the O *K*-edge. The hybridization between the O 2*p* and the Rh 4*d* as indicated by the partial DOS (see Fig. 3) enable *d* states to be investigated here. The three peaks shown in Fig. 5b are due to the Rh 4*d* manifold. The small peak at ~528.5 eV is $t_{2g}$ in character while the others at ~530.5 and ~ 532.5 eV are $e_g$ in character. This interpretation is in great agreement with the line shapes of the partial DOS (Fig. 3b), where the states near $E_F$ have strong contributions from both Rh $d$ and O $p$ orbitals, whereas there is a strong peak ~7 eV above $E_F$ that is mostly Sr $d$ in character. Making comparison to $SrCoO_3$[28] and $SrIrO_3$ that are respectively the 3$d$ and 5$d$ analogs of $SrRhO_3$, we find that peaks associated with $d$ orbitals are much broader for $SrRhO_3$, implying that hybridization of the O $p$ and Rh $d$ manifolds is likely quite strong here. Combining this with the fact that recent experimental studies have revealed that both $SrCoO_3$[29] and $SrIrO_3$[30] have significantly improved catalytic activity in oxygen evolution reactions as compared to Pt and $IrO_2$, respectively and that for the former this behavior is accompanied by such hybridization implies that $SrRhO_3$ has great promise for enhanced oxygen kinetics while such investigations are part of our continued work that is beyond the scope of this study.

The electronic properties were further investigated by SE (Fig. 5c) and for $SrRhO_3$ we observed a nearly linear response, however if one considers the calculated DOS in this energy range both above and below $E_F$ it is approximately constant which should yield a nearly featureless optical conductivity signal. Despite this, there does appear to be a weak broad peak at roughly 2 eV. Comparing this spectra to that of $SrIrO_3$ ($SrCoO_3$),[31] we observe a much sharper peak at 3.3 eV (1.8 eV) that is due to *p-d* optical transitions. Thus, there is a systematic shift of this feature to higher energy with increasing atomic number and similar to XAS data, it is much broader than in either counterpart. Therefore, there is great agreement between both spectroscopic techniques and our DFT calculations.

In summary, we have successfully synthesized high quality epitaxial films of $SrRhO_3$ by PLE and have investigated their electronic and magnetic properties through various experimental and theoretical techniques. We conclude that the magnetic ground state in $SrRhO_3$ is highly sensitive to rotations of the $RhO_6$ octahedra and C-type AFM is the energetically favorable state at zero temperature, which drastically differs from that of bulk crystals. It has an ordering temperature of roughly 110 K and likely originates from substrate induced high symmetry phase where the octahedra are not rotated ($a^0a^0a^0$), thus providing an example where the octahedral distortions are tunable and strongly coupled to physical properties. We have made a number of comparisons between these experimental and theoretical results and find them to be highly self-consistent. In addition, these SrRhO$_3$ films exhibit abnormally strong hybridization between the O 2$p$ and Rh 4$d$. Thus, SrRhO$_3$ displays great potential for novel functionalities and synthesis of high quality single crystal samples is no longer a hurdle for experimentalists.

## Acknowledgements

This work was supported by the US Department of Energy (DOE), Office of Science, Basic Energy Sciences (BES), Materials Sciences and Engineering Division (experiment) and through the Office of Science Early Career Research Program (theory). All computations were performed at the National Energy Research Scientific Computing Center (NERSC), which is supported by the Office of Science of the U.S. DOE under Contract No. DE-AC02-05CH11231. Optical measurements were conducted at the Center for Nanophase Materials Sciences, which is a DOE Office of Science User Facility. Use of the Advanced Photon Source, an Office of Science User Facility operated for the US DOE, Office of Science by Argonne National Laboratory, was supported by the US DOE under contract no. DE-AC02- 06CH11357 (XAS).

# References

[1] E. Dagotto, *Science* **309**, 257 (2005).
[2] Y. Tokura and N. Nagaosa, *Science* **288**, 462 (2000).
[3] Y. Maeno, H. Hashimoto, K. Yoshida, S. Nishizaki, T. Fujita, J. G. Bednorz, and F. Lichtenberg, *Nature* **372**, 532 (1994).
[4] K. Yamaura and E. Takayama-Muromachi, *Physica C* **445–448**, 54 (2006).
[5] K. Yamaura and E. Takayama-Muromachi, *Phys. Rev. B* **64**, 224424 (2001).
[6] K. Yamaura, Q. Huang, D. P. Young, Y. Noguchi, and E. Takayama-Muromachi, *Phys. Rev. B* **66**, 134431 (2002).
[7] K. Yamaura, Q. Huang, D. P. Young, and E. Takayama-Muromachi, *Chemistry of Materials* **16**, 3424 (2004).
[8] K. Yamaura, Y. Shirako, H. Kojitani, M. Arai, D. P. Young, M. Akaogi, M. Nakashima, T. Katsumata, Y. Inaguma, and E. Takayama-Muromachi, *J. Am. Chem. Soc.* **131**, 2722 (2009).
[9] Y. Shirako, H. Kojitani, M. Akaogi, K. Yamaura, and E. Takayama-Muromachi, *Physics and Chemistry of Minerals* **36**, 455 (2009).
[10] C.-Q. Jin, J.-S. Zhou, J. B. Goodenough, Q. Q. Liu, J. G. Zhao, L. X. Yang, Y. Yu, R. C. Yu, T. Katsura, A. Shatskiy, and E. Ito, *Proceedings of the National Academy of Sciences* **105**, 7115 (2008).
[11] D. J. Singh, *Phys. Rev. B* **67**, 054507 (2003).
[12] J.-M. Carter, V. V. Shankar, M. A. Zeb, and H.-Y. Kee, *Phys. Rev. B* **85**, 115105 (2012).
[13] D. Xiao, W. Zhu, Y. Ran, N. Nagaosa, and S. Okamoto, *Nat Commun* **2**, 596 (2011).
[14] Y. X. Liu, H. Masumoto, and T. Goto, *Materials Transactions* **46**, 100 (2005).
[15] S. Havelia, K. R. Balasubramaniam, S. Spurgeon, F. Cormack, and P. A. Salvador, *Journal of Crystal Growth* **310**, 1985 (2008).
[16] J. Nichols, O. B. Korneta, J. Terzic, G. Cao, J. W. Brill, and S. S. A. Seo, *Appl. Phys. Lett.* **104**, 121913 (2014).
[17] N. F. Mott, *Philos. Mag.* **26**, 1015 (1972).
[18] A. F. Ioffe and A. R. Regel, *Prog. Semicond.* **4**, 237 (1960).
[19] N. E. Hussey ∥, K. Takenaka, and H. Takagi, *Philos. Mag.* **84**, 2847 (2004).
[20] M. Takizawa, D. Toyota, H. Wadati, A. Chikamatsu, H. Kumigashira, A. Fujimori, M. Oshima, Z. Fang, M. Lippmaa, M. Kawasaki, and H. Koinuma, *Phys. Rev. B* **72**, 060404 (2005).
[21] Y. J. Chang, C. H. Kim, S. H. Phark, Y. S. Kim, J. Yu, and T. W. Noh, *Phys. Rev. Lett.* **103**, 057201 (2009).
[22] B. Kim, D. Kwon, J. H. Song, Y. Hikita, B. G. Kim, and H. Y. Hwang, *Solid State Commun.* **150**, 598 (2010).
[23] J. Son, J. M. LeBeau, S. J. Allen, and S. Stemmer, *Appl. Phys. Lett.* **97**, 202109 (2010).
[24] T. L. Meyer, L. Jiang, S. Park, T. Egami, and H. N. Lee, *APL Mater.* **3**, 126102 (2015).
[25] J. M. Rondinelli, S. J. May, and J. W. Freeland, *MRS Bulletin* **37**, 261 (2012).
[26] Z. Liao, M. Huijben, Z. Zhong, N. Gauquelin, S. Macke, R. J. Green, S. Van Aert, J. Verbeeck, G. Van Tendeloo, K. Held, G. A. Sawatzky, G. Koster, and G. Rijnders, *Nat Mater* **15**, 425 (2016).
[27] N. Han-Jin, B. J. Kim, S. J. Oh, J. H. Park, H. J. Lin, C. T. Chen, Y. S. Lee, K. Yamaura, and E. Takayama-Muromachi, *Journal of Physics: Condensed Matter* **20**, 485208 (2008).
[28] H. Jeen, W. S. Choi, M. D. Biegalski, C. M. Folkman, I. C. Tung, D. D. Fong, J. W. Freeland, D. Shin, H. Ohta, M. F. Chisholm, and H. N. Lee, *Nat Mater* **12**, 1057 (2013).
[29] J. R. Petrie, C. Mitra, H. Jeen, W. S. Choi, T. L. Meyer, F. A. Reboredo, J. W. Freeland, G. Eres, and H. N. Lee, *Advanced Functional Materials* **26**, 1564 (2016).
[30] R. Tang, Y. Nie, J. K. Kawasaki, D.-Y. Kuo, G. Petretto, G. Hautier, G.-M. Rignanese, K. M. Shen, D. G. Schlom, and J. Suntivich, *J. Mater. Chem. A* **4**, 6831 (2016).

31 W. S. Choi, H. Jeen, J. H. Lee, S. S. A. Seo, V. R. Cooper, K. M. Rabe, and H. N. Lee, *Phys. Rev. Lett.* **111**, 097401 (2013).

Figure Captions

**Figure 1** AFM topographic images of 3 × 3 μm$^2$ area of the STO substrate obtained (a) before and (b) after deposition of a ~12 nm thick SrRhO3 film. (c) XRD $\theta$-$2\theta$ scan of SrRhO$_3$ film..

**Figure 2** (a) The temperature dependence of the longitudinal resistivity ($\rho_{xx}$). (inset) The field dependence of the transverse resistivity ($\rho_{xy}$) obtained at $T$ = 2 K. (b) The temperature dependent magnetoresistance ($MR = [R(H) - R(0)]/R(0) \times 100\%$) obtained at $\mu_\circ H$ = 5 T. (inset) The field dependent $MR$ obtained at $T$ = 2 K. (c) The temperature dependent magnetization ($M$) obtained while warming with $\mu_\circ H$ = 0.1 T after cooling in the same and zero field.

**Figure 3** Calculated total and partial DOS of SrRhO$_3$ at $T$ = 0K for (a) C-type AF and (b) PM magnetic ground states obtained with an ideal thin film crystal structure with undistorted (a$^0$a$^0$a$^0$) RhO$_6$ octahedra (space group *Amm2*). The dashed gray line marks DOS = 0 and the inset shows a magnification near $E_F$.

**Figure 4** X-ray diffractions reciprocal space maps near 103, 013, 103, and 013 reflections of STO.

**Figure 5** (a) Room temperature XAS spectra near the O *K* edge of both SrRhO$_3$ and SrTiO$_3$ where there is a vertical offset for clarity. (b) Room temperature XAS spectra (TEY) near the O *K* edge of SrIrO$_3$, SrRhO$_3$, and SrCoO$_3$. (c) Optical conductivity ($\sigma_1$) of SrRhO$_3$ film obtained at room temperature along with SrCoO$_3$ (from Ref. 31) and SrIrO$_3$.

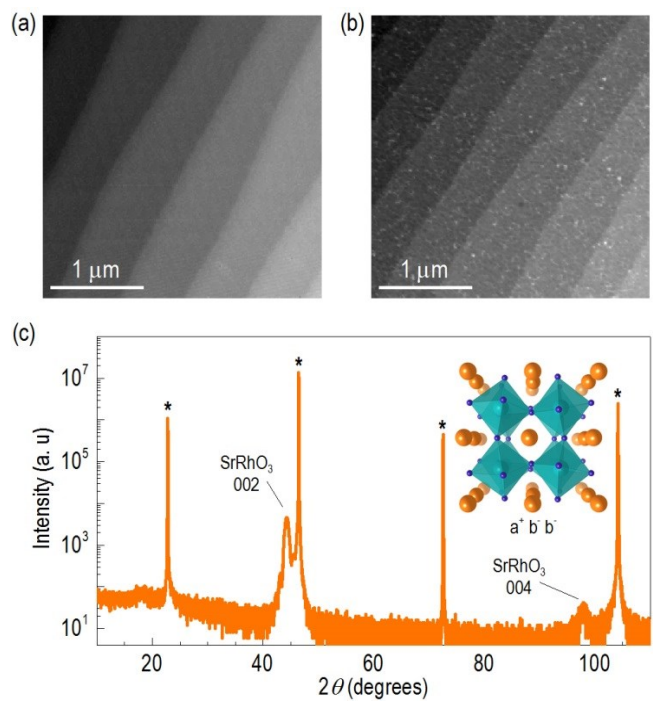

Nichols *et al*.
Figure 1

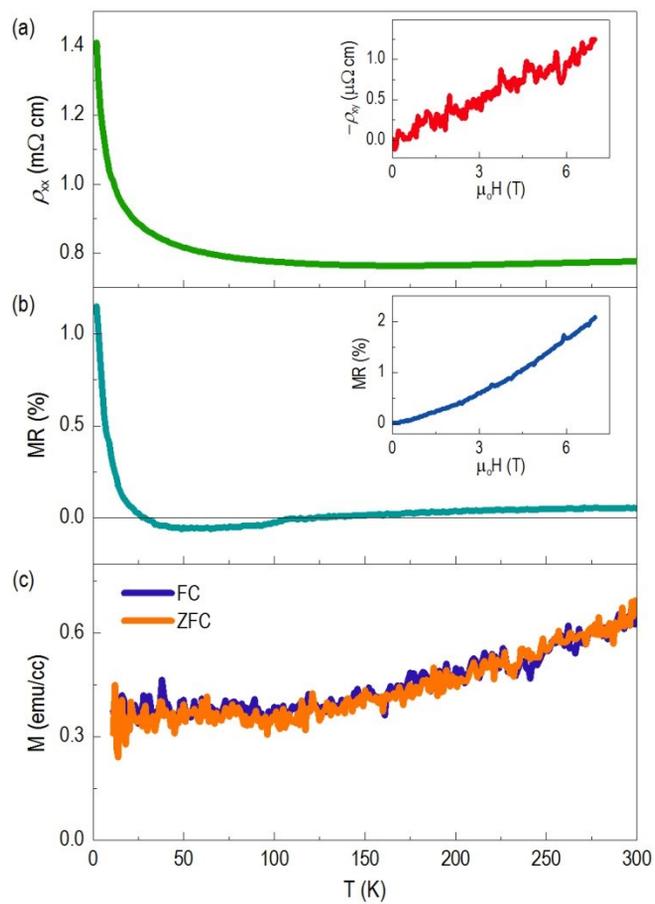

Nichols *et al*.
Figure 2

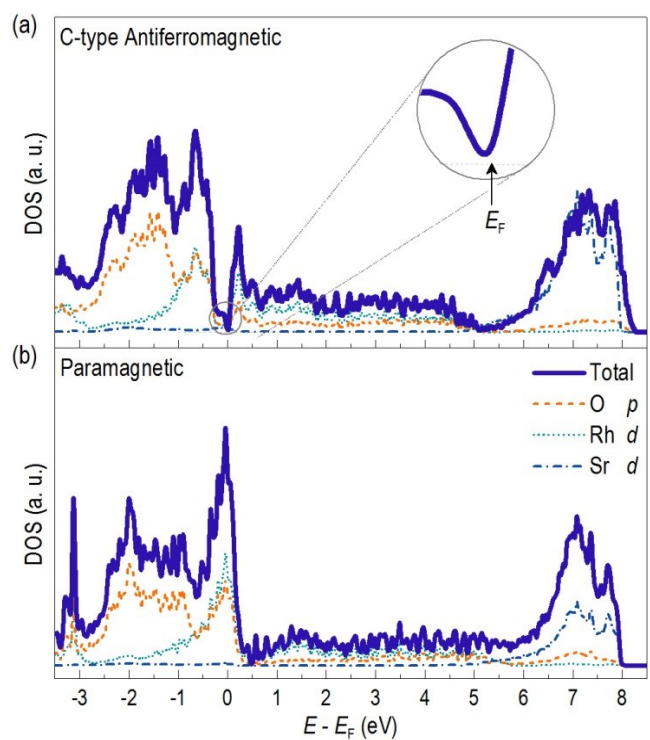

Nichols *et al*.
Figure 3

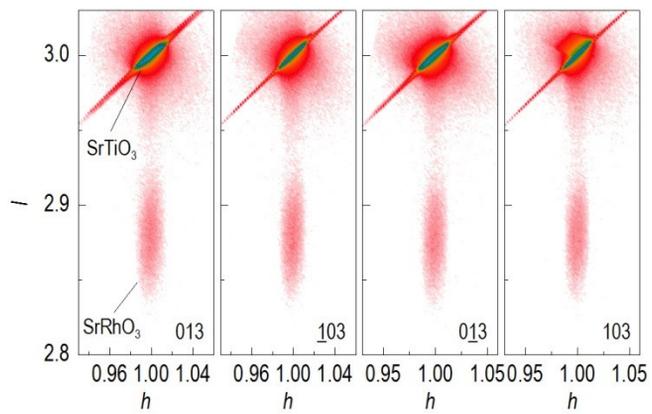

Nichols *et al.*
Figure 4

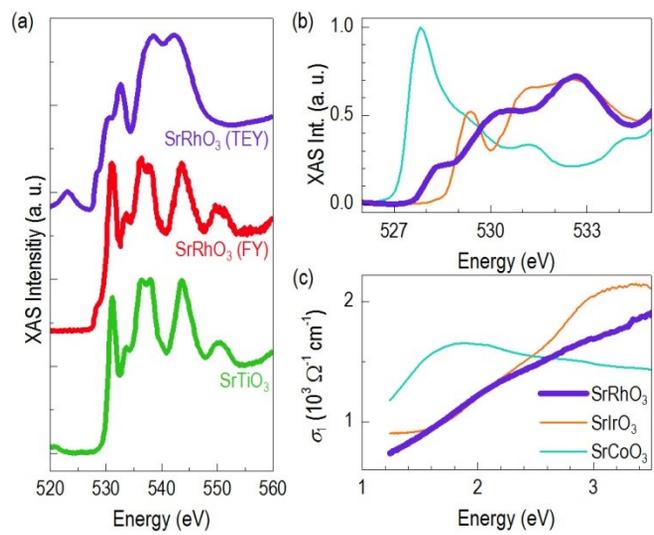

Nichols *et al.*
Figure 5